\documentclass[preprint,aps]{revtex4}

\usepackage{graphicx}
\usepackage{dcolumn}
\usepackage{bm}

\begin{document}

\title{Squeezed States of the Generalized Minimum Uncertainty State
for the Caldirola-Kanai Hamiltonian}

\author{Sang Pyo Kim}\email{sangkim@kunsan.ac.kr}

\affiliation{Department of Physics, Kunsan National University,
Kunsan 573-701, Korea}

\affiliation{Asia Pacific Center for Theoretical Physics, Pohang
790-784, Korea}

\date{\today}
\begin{abstract}
We show that the ground state of the well-known pseudo-stationary
states for the Caldirola-Kanai Hamiltonian is a generalized
minimum uncertainty state, which has the minimum allowed
uncertainty $\Delta q \Delta p = \hbar \sigma_0/2$, where
$\sigma_0 (\geq 1)$ is a constant depending on the damping factor
and natural frequency. The most general symmetric Gaussian states
are obtained as the one-parameter squeezed states of the
pseudo-stationary ground state. It is further shown that the
coherent states of the pseudo-stationary ground state constitute
another class of the generalized minimum uncertainty states.
\end{abstract}

\maketitle

\section{Introduction}

The Hamiltonian for a harmonic oscillator with an exponentially
increasing mass has been introduced by Caldirola and Kanai
\cite{cal} and the corresponding Lagrangian by Bateman \cite{bat}.
The fact that its classical motion describe a damping motion has
motivated the investigation of the Caldirola-Kanai (CK)
Hamiltonian as a quantum damped system \cite{dek}. The
pseudo-stationary states of the CK Hamiltonian have been found in
many different ways
\cite{ker,bop,has,dod,jan,che,cer,um,che2,bri,sri,ali, kim,ped}.
In particular, the invariant operator method provides a convenient
tool to find exact wave functions for such time-dependent
oscillators \cite{lew}. However, there have been debates whether
this quantum oscillator genuinely describes a dissipative system
or not \cite{brittin,ker,has,dod}.

In this paper we show that all the Gaussian states of the CK
Hamiltonian with $\langle \hat{q} \rangle = 0 = \langle \hat{p}
\rangle$ satisfy the generalized minimum uncertainty relation
\begin{equation}
\Delta q \Delta p \geq \frac{\hbar}{2} \sigma_0, \quad (\sigma_0
\geq 1), \label{gmu}
\end{equation}
where $\sigma_0 = 1/(1 - \gamma^2/4\omega_0^2)^{1/2}$ is a
constant depending on the damping factor $\gamma$ and the natural
frequency $\omega_0$. It is shown that the pseudo-stationary
ground state is the generalized minimum uncertainty state (GMUS),
a generalization of the minimum uncertainty state with $\sigma_0 =
1$ \cite{sto}. Using the linear invariant operators
\cite{dod2,kim2,kim-page}, we find the most general Gaussian
states for the CK Hamiltonian, which have the zero moment of
position and momentum, and show that the pseudo-stationary ground
state is, in fact, a GMUS. The GMUS that is symmetric about the
origin is interpreted as the vacuum state of time-dependent
oscillator in Ref. \cite{kim2}. We further show that the coherent
states of the pseudo-stationary ground state are also the GMUS's.

\section{Squeezed States of Pseudo-Stationary States}

The harmonic oscillator with an exponentially increasing mass $m =
m_0 e^{\gamma t}$ has the CK Hamiltonian
\begin{equation}
\hat{H} (t) = \frac{1}{2m_0} e^{- \gamma t} \hat{p}^2 + \frac{m_0
\omega_0^2}{2}
 e^{\gamma t} \hat{q}^2. \label{osc}
\end{equation}
The Hamilton equations describe a classical damped motion
\begin{equation}
\ddot{u} + \gamma \dot{u} + \omega_0^2 u = 0. \label{cl eq}
\end{equation}
Now we use the invariant operator method to find exact quantum
states of the time-dependent CK Hamiltonian. For each complex
solution $u$ of Eq. (\ref{cl eq}), one can introduce a pair of
linear invariant operators \cite{kim-page}
\begin{eqnarray}
\hat{a} (t)  &=& \frac{i}{\sqrt{\hbar}} [u^*(t) \hat{p}
- m_0 e^{\gamma t} \dot{u}^*(t)  \hat{q} ], \nonumber\\
\hat{a}^{\dagger} (t) &=& - \frac{i}{\sqrt{\hbar}} [u (t) \hat{p}
- m_0 e^{\gamma t} \dot{u} (t) \hat{q}]. \label{inv op}
\end{eqnarray}
In fact, these operators can be made the time-dependent
annihilation and creation operators satisfying the standard
commutation relation at equal time
\begin{equation}
[\hat{a} (t), \hat{a}^{\dagger} (t) ] = 1,
\end{equation}
by imposing the Wronskian condition
\begin{equation}
m_0 e^{\gamma t} [u (t) \dot{u}^* (t) - u^* (t) \dot{u}(t)] = i.
\label{wron}
\end{equation}
We note that the eigenfunctions of $\hat{a}^{\dagger} (t) \hat{a}
(t)$, another invariant operator, \cite{kim-page}
\begin{eqnarray}
\Psi_n (q, t) = \Biggl(\frac{1}{2^{n} n! \sqrt{2 \pi \hbar u^* u}}
\Biggr)^{1/2} \Biggl(\frac{u}{\sqrt{u^* u}} \Biggr)^{n + 1/2} H_n
\Biggl(\frac{q}{\sqrt{2 \hbar u^* u}} \Biggr) \exp \Biggl[\frac{i
m_0 e^{\gamma t} \dot{u}^*}{2 \hbar u^*} q^2\Biggr] \label{ex
wave}
\end{eqnarray}
are the exact quantum state of the Schr\"{o}dinger equation
\cite{lew}. Hence the task to find the general wave functions is
equivalent to finding the general solutions to Eq. (\ref{cl eq}).
Our stratagem is to select a complex solution $u_0$ satisfying Eq.
(\ref{wron}) and, as Eq. (\ref{cl eq}) is linear, to find the
general solution as a linear superposition of $u_0$ and $u^*_0$.

For the underdamped motion $(\omega_0 > \gamma/2)$, we select the
solution
\begin{equation}
u_0 (t) = \frac{e^{- \gamma t/2}}{\sqrt{2 m_0 \omega}} ~ e^{- i
\omega t}, \quad \omega = \sqrt{\omega_0^2 - \frac{\gamma^2}{4}}.
\label{min sol}
\end{equation}
Then the wave functions of number states with the solution
(\ref{min sol}) substituted into Eq. (\ref{ex wave}) yield the
pseudo-stationary states
\cite{ker,bop,has,dod,jan,che,cer,um,che2,bri,sri,ali, kim,ped}
\begin{eqnarray}
\Psi_n (q, t) = \frac{1}{\sqrt{2^n n!}} \Biggl(\frac{m_0 \omega
e^{\gamma t} }{ \pi \hbar} \Biggr)^{1/4} e^{- i \omega t (n +
1/2)} H_n \Biggl(\sqrt{\frac{m_0 \omega}{\hbar} e^{\gamma t}} q
\Biggr) \exp \Biggl[ - \frac{m_0 \omega e^{\gamma t}}{2 \hbar}
\Biggl(1 + i \frac{\gamma}{2 \omega} \Biggr) q^2 \Biggr].
\label{min wave}
\end{eqnarray}
Now, the general complex solutions satisfying the quantization
condition (\ref{wron}) are written as
\begin{equation}
u_r (t) = \mu u_0 (t) + \nu u_0^* (t),
\end{equation}
where
\begin{equation}
|\mu|^2 - |\nu|^2 = 1. \label{con}
\end{equation}
The complex $\mu$ and $\nu$ have four real parameters, one of
which is constrained by Eq. (\ref{con}), and the other of which
can be absorbed into the overall phase of $u_r$ and hence does not
change the wave functions. However, the relative phase between
$\mu$ and $\nu$ is not determined by constraints. The squeezing
parameters $r$ and $\phi$ in the form
\begin{equation}
\mu = \cosh r, \quad \nu = e^{i \phi} \sinh r,
\end{equation}
are, in fact, two integration constants of the second order
equation (\ref{cl eq}). Conversely, given any complex solution $u$
satisfying Eq. (\ref{wron}), we can find the corresponding
parameters $\mu$ and $\nu$ or $r$ and $\phi$. Therefore, the most
general solution to Eq. (\ref{cl eq}) can be written as
\begin{equation}
u_{r \phi} (t) = (\cosh r) u_0 (t) + (e^{i \phi} \sinh r) u_0^*
(t). \label{gen sol}
\end{equation}
That $r$ and $\phi$ are the squeezed parameters is understood from
the Bogoliubov transformation
\begin{eqnarray}
\hat{a}_{r \phi} (t) &=& \mu^* \hat{a}_0 (t) - \nu^*
\hat{a}^{\dagger}_0 (t),
\nonumber\\
\hat{a}^{\dagger}_{r \phi} (t) &=& \mu \hat{a}^{\dagger}_0 (t) -
\nu \hat{a}_0 (t),
\end{eqnarray}
which is obtained by substituting Eq. (\ref{gen sol}) into Eq.
(\ref{inv op}). The Bogoliubov transformation is a unitary
transformation of $\hat{a}_0 (t)$ and $\hat{a}_0^{\dagger} (t)$:
\begin{eqnarray}
\hat{a}_{r \phi} (t) &=& \hat{U}(z, t) \hat{a}_0 (t)
\hat{U}^{\dagger}(z,t), \nonumber\\
\hat{a}_{r \phi}^{\dagger} (t) &=& \hat{U}(z, t)
\hat{a}_0^{\dagger} (t) \hat{U}^{\dagger}(z,t),
\end{eqnarray}
where
\begin{equation}
\hat{U} (t, z) = \exp \Biggl[\frac{1}{2} \Biggl( z
\hat{a}^{\dagger 2}_0 (t) - z^* \hat{a}^2_0(t) \Biggr) \Biggr],
\quad z = e^{i (\phi+ \pi)} r,
\end{equation}
is the squeeze operator \cite{sto}.

Each pair of squeeze parameters $r$ and $\phi$ defines a family of
the invariant number operators
\begin{equation}
\hat{N}_{r \phi} (t) = \hat{a}^{\dagger}_{r \phi} (t) \hat{a}_{r
\phi} (t).
\end{equation}
The number states
\begin{equation}
\hat{N}_{r \phi} (t) \vert n, r, \phi, t \rangle = n \vert n, r,
\phi, t \rangle
\end{equation}
lead to the exact wave functions (\ref{ex wave}) for the
Schr\"{o}dinger equation in the form
\begin{equation}
\Psi_n (q, t, r, \phi) = \frac{1}{\sqrt{2^n n!}} \Biggl(
\frac{A_{r \phi}}{\sqrt{\pi}} \Biggr)^{1/2} e^{- i \Theta_{r \phi}
(n + 1/2)} H_n (A_{r \phi} q) e^{- B_{r \phi} q^2}, \label{gen
wave}
\end{equation}
where
\begin{eqnarray}
A_{r \phi} &=& \frac{1}{\sqrt{2 \hbar u_{r \phi}^* u_{r \phi}}} =
\sqrt{\frac{m_0 \omega e^{\gamma t} }{\hbar}} \frac{1}{[\cosh 2r +
\sinh 2r \cos (2 \omega t + \phi)]^{1/2}},
\nonumber\\
B_{r \phi} &=& \frac{i m \dot{u}_{r \phi}^*}{2 \hbar u_{r \phi}^*}
= \frac{m_0 \omega e^{\gamma t}}{2 \hbar} \Biggl[\frac{\cosh r
e^{i \omega t} - e^{- i \phi} \sinh r e^{- i \omega t} }{\cosh r
e^{i \omega t} + e^{- i \phi} \sinh r e^{- i \omega t}} + i
\frac{\gamma}{2 \omega} \Biggr], \nonumber\\
\Theta_{r \phi} &=& \tan^{-1} \Biggl[ \frac{\sin \omega t - \tanh
r \sin (\omega t + \phi)}{\cos \omega t + \tanh r \cos (\omega t +
\phi)} \Biggr]. \label{coeff}
\end{eqnarray}
Here $\Theta_{r \phi}$ is the negative phase of $u_{r \phi}$, that
is, $u_{r \phi} = \rho_r e^{-i \Theta_{r \phi}}$. The wave
functions (\ref{gen wave}), which are symmetric about the origin
($\langle \hat{q} \rangle = \langle \hat{p} \rangle = 0$), are the
squeezed states of the pseudo-stationary states (\ref{min wave}).
Besides the zero squeezing parameter $(r = 0)$ leading to the
pseudo-stationary states, another interesting squeezing parameters
\begin{eqnarray}
\cosh 2 r_0 = 1 + \frac{\gamma^2}{8 \omega^2}, \quad \tan \phi_0 =
\frac{4 \omega}{\gamma}
\end{eqnarray}
lead to the simple harmonic wave functions at $t = 0$:
\begin{equation}
\Psi_n (q, t = 0, r_0, \phi_0) = \exp \Biggl[ - i \frac{\gamma}{4
\omega} \Biggl( n + \frac{1}{2} \Biggr) \Biggr]  \times \Biggl\{
\frac{1}{\sqrt{2^n n!}} \Biggl(\frac{m_0 \omega}{\pi \hbar}
\Biggr)^{1/4}H_n \Biggl(\sqrt{\frac{m_0 \omega}{\hbar}} q \Biggr)
\exp \Biggl[ - \frac{m_0 \omega}{2 \hbar} q^2 \Biggr] \Biggr\}.
\label{sim wave}
\end{equation}
The wave functions (\ref{gen wave}), evolving the harmonic wave
functions of an undamped $(\gamma = 0)$ oscillator at $t = 0$,
differ from those in Ref. \cite{um} only by the constant phase
factor in Eq. (\ref{sim wave}).

\section{Generalized Minimum Uncertainty State}

We now find the GMUS satisfying the equality in Eq. (\ref{gmu})
among the wave functions (\ref{gen wave}), which are symmetric
about the origin. The wave functions (\ref{gen wave}) have the
uncertainty
\begin{eqnarray}
(\Delta q)_{n r \phi} (\Delta p)_{n r \phi} &=& \langle n,r, \phi,
t \vert \hat{q}^2 \vert n, r, \phi, t \rangle^{1/2}
\langle n, r, \phi, t \vert \hat{p}^2
\vert n, r, \phi, t \rangle^{1/2} \nonumber\\
&=& \frac{\hbar}{2}
\sec\Biggl(\frac{\vartheta_{\gamma}}{2}\Biggr)[\{ \cosh 2r + \sinh
2 r \cos(2 \omega t + \phi)\}\nonumber\\&& \times \{\cosh 2r -
\sinh 2r \cos(2 \omega t + \phi + \vartheta_{\gamma})\}]^{1/2}
\Biggl(n + \frac{1}{2}\Biggr), \label{gen unc}
\end{eqnarray}
where
\begin{equation}
\vartheta_{\gamma} = \sin^{-1}
\Biggl(\frac{\frac{\gamma}{\omega}}{1 + \frac{\gamma^2}{4 \omega^2
}} \Biggr) = \cos^{-1} \Biggl(\frac{1 - \frac{\gamma^2}{4 \omega^2
}}{1 + \frac{\gamma^2}{4 \omega^2 }} \Biggr), \quad (\pi >
\vartheta_{\gamma} \geq 0).
\end{equation}
Using Eq. (\ref{gen unc}, we find the condition leading to the
minimum allowed uncertainty. First, from $(\Delta q)_{nr\phi}
(\Delta p)_{nr\phi} = (\Delta q)_{0 r \phi} (\Delta p)_{0 r \phi}
(n + 1/2)$, the ground state $(n = 0)$ has the lower uncertainty
than other excited states $(n \geq 1)$. Second, for the zero
squeezing parameter $(r = 0)$, the pseudo-stationary ground state
$\Psi_0 (q, t)$ has the generalized minimum uncertainty at all
times
\begin{eqnarray}
(\Delta q)_{00 \phi} (\Delta p)_{00 \phi} &=& \frac{\hbar}{2}
\sec\Biggl(\frac{\vartheta_{\gamma}}{2}\Biggr).
\end{eqnarray}
Thus the generalized minimum uncertainty (\ref{gmu}) is satisfied
for
\begin{equation}
\sigma_0 = \sec \Biggl( \frac{\vartheta_{\gamma}}{2} \Biggr) =
\frac{1}{\Biggl( 1 - \frac{\gamma^2}{4 \omega^2_0} \Biggr)^{1/2}}.
\end{equation}
Note that the generalized minimum uncertainty approaches the usual
minimum uncertainty $(\hbar/2)$ in the weak damping limit
$(\gamma/\omega_0 \ll 1)$. Similarly the time averaged uncertainty
is
\begin{eqnarray}
\overline{(\Delta q)_{0 r \phi} (\Delta p)_{0 r \phi}} &=&
\frac{\hbar}{2} \sec\Biggl(\frac{\vartheta_{\gamma}}{2}\Biggr)
\Biggl( \cosh^2 r - \frac{\cos \vartheta_{\gamma}}{2}\sinh^2 r
\Biggr)
\nonumber\\
&\geq& \frac{\hbar}{2}
\sec\Biggl(\frac{\vartheta_{\gamma}}{2}\Biggr),
\end{eqnarray}
where the equality holds for $r = 0$. Third, in the case of zero
damping $(\gamma = 0 = \vartheta_{\gamma})$, the CK Hamiltonian
(\ref{osc}) is just a simple (time-independent) harmonic
oscillator. Then the uncertainty relation of $\hat{q}$ and
$\hat{p}$ in the state (\ref{gen wave}) is given by
\begin{eqnarray}
(\Delta q)_{0 r \phi} (\Delta p)_{0 r \phi} &=& \frac{\hbar}{2} [
\cosh^2 (2r) - \sinh^2 (2 r) \cos^2(2 \omega t + \phi)]^{1/2} \nonumber\\
&\geq& \frac{\hbar}{2}.
\end{eqnarray}
The generalized minimum uncertainty is achieved either for the
zero squeezing $(r = 0)$ at all times or when $\cos (2 \omega t +
\phi) = \pm 1$. Therefore, we conclude that the pseudo-stationary
ground state, which is provided by the zero squeezing $(r = 0)$
solution $u_0$ in Eq. (\ref{min sol}), gives rise to the GMUS with
the center at the origin. In particular, this GMUS is interpreted
as the vacuum state in Ref. \cite{kim2}. Finally we obtain the
Hamiltonian expectation value
\begin{equation}
\langle \hat{H} \rangle_{n r \phi} = \frac{\hbar \omega}{2} \sec^2
\Biggl(\frac{\vartheta_{\gamma}}{2}\Biggr) \Biggl[ \cosh 2r +
\sinh 2r \sin\Biggl( \frac{\vartheta_{\gamma}}{2} \Biggr) \sin
\Biggl(2 \omega t + \phi + \frac{\vartheta_{\gamma}}{2} \Biggr)
\Biggr] \Biggl(n + \frac{1}{2} \Biggr).
\end{equation}
The time averaged $\overline{\langle \hat{H} \rangle}_{n r \phi}$
has the minimum value for $n = r = 0$, coinciding with the
generalized minimum uncertainty.

There is another class of GMUS's. It is known that for a
time-independent oscillator, the coherent states of the vacuum
state also have the minimum uncertainty \cite{sto}. Now, for the
CK Hamiltonian, we either follow the definition of coherent states
\cite{har,raj}
\begin{equation}
\hat{a}_{r \phi} (t) \vert \alpha, r, \phi, t \rangle = \alpha
\vert \alpha, r, \phi, t \rangle,
\end{equation}
for any complex $\alpha$ or apply the displacement operator to the
ground state in Eq. (\ref{gen wave})
\begin{equation}
\vert \alpha, r, \phi, t \rangle = e^{\alpha \hat{a}^{\dagger}_{r
\phi} (t) - \alpha^* \hat{a}_{r \phi} (t)} \vert 0, r, \phi, t
\rangle.
\end{equation}
Then the generalized coherent states have the expectation values
\begin{eqnarray}
q_c (t) &=& \langle \alpha, r, \phi, t \vert \hat{q} \vert \alpha,
r, \phi, t \rangle = \sqrt{\hbar} (\alpha u_{r \phi}
+ \alpha^* u^*_{r \phi}), \nonumber\\
p_c (t) &=& \langle \alpha, r, \phi, t \vert \hat{p} \vert \alpha,
r, \phi, t \rangle = \sqrt{\hbar} m_0 e^{\gamma t} (\alpha
\dot{u}_{r \phi} + \alpha^* \dot{u}^*_{r \phi}).
\end{eqnarray}
Here $q_c$ and $p_c$ describe a trajectory in the phase space for
each choice of $\alpha$ and $u_{r \phi}$. Replacing the complex
$\alpha$ by two real variables $q_c$ and $p_c$, we obtain the wave
functions for the coherent states
\begin{equation}
\Psi (q, t, r, \phi, q_c, p_c) = \Biggl(\frac{A_{r
\phi}}{\sqrt{\pi}} \Biggr)^{1/2} F_{r \phi} e^{ - i \Theta_{r
\phi} /2} e^{- B_{r \phi} (q - q_c)^2} e^{i p_c q/ \hbar},
\label{coh}
\end{equation}
where $A_{r \phi}$, $B_{r \phi}$, and $\Theta_{r \phi}$ are given
in Eq. (\ref{coeff}) and $F_{r \phi}$ is the additional phase
factor
\begin{equation}
F_{r \phi} = \exp \Biggl[ \frac{i}{2 \hbar \dot{u}_{r\phi}^* u_{r
\phi} ^* } (u_{r \phi}^{*2} p_c^2 - 2 \dot{u}_{r \phi}^* u^*_{r
\phi} p_c q_c) \Biggr].
\end{equation}
It then follows that any coherent state (\ref{coh}) has the same
uncertainty as the general Gaussian state with $n = 0$ in Eq.
(\ref{gen wave}):
\begin{eqnarray}
(\Delta q)_{\alpha r \phi} (\Delta p)_{\alpha r \phi} &=& \langle
\alpha,r, \phi, t \vert (\hat{q} - \langle \hat{q} \rangle)^2
\vert \alpha, r, \phi, t \rangle^{1/2} \langle \alpha,r, \phi, t
\vert (\hat{p} - \langle \hat{p}
\rangle)^2 \vert \alpha, r, \phi, t \rangle^{1/2} \nonumber\\
&=& (\Delta q)_{0 r \phi} (\Delta p)_{0 r \phi}.
\end{eqnarray}
This implies that the coherent states of the GMUS are also GMUS's.
Thus the coherent states of the pseudo-stationary ground state
constitute a family of GMUS's, which is the time-dependent
generalization of time-independent oscillator \cite{sto}.

\section{Conclusion}

We have shown that the Caldirola-Kanai Hamiltonian satisfies the
generalized minimum uncertainty $\Delta q \Delta p \geq \hbar
\sigma_0/2$ for $\sigma_0 = 1/ (1 - \gamma^2/4 \omega_0^2)^{1/2}$,
where $\gamma$ is the damping factor and $\omega_0$ is the natural
frequency. It is found that the well-known pseudo-stationary
ground state has in fact the generalized minimum uncertainty. As
the generalized minimum uncertainty state is uniquely selected for
the Caldirola-Kanai Hamiltonian, this pseudo-stationary ground
state may be interpreted as the vacuum state \cite{kim2}.
One-parameter family of squeezed states of the pseudo-stationary
states are obtained as the most general states with the zero
moment of position and moment. Further, it is shown that the
coherent states of the pseudo-stationary ground state are the
generalized minimum uncertainty states.

\acknowledgements

The author would like to thank J.Y. Kim and C.-I. Um for useful
discussions. This work was supported by the Korea Research
Foundation under Grant No. KRF-2002-041-C00053.

\end{document}